\documentclass[conference]{IEEEtran}
\usepackage{cite}
\usepackage{amsmath,amssymb,amsfonts}
\usepackage{algorithmic}
\usepackage{graphicx}
\usepackage{textcomp}
\usepackage{xcolor}
\usepackage{booktabs}
\usepackage{multirow}
\usepackage{hyperref}
\usepackage{subcaption}

\def\BibTeX{{\rm B\kern-.05em{\sc i\kern-.025em b}\kern-.08em
    T\kern-.1667em\lower.7ex\hbox{E}\kern-.125emX}}

\begin{document}

\title{Risk-Aware GPU-Assisted Cardinality Estimation for Cost-Based Query Optimizers}

\author{\IEEEauthorblockN{Il-Sun Chang}
\IEEEauthorblockA{\textit{Independent Researcher} \\
Daegu, Republic of Korea \\
kyou0072@gmail.com}
}

\maketitle

\begin{abstract}
Cardinality estimation is the cornerstone of Cost-Based Optimizers (CBO), yet we empirically \textbf{measure and quantify} how statistical failures degrade \textbf{decision stability} and increase \textbf{plan flip rates}. By analyzing the \textbf{overhead breakdown and break-even} points of hardware acceleration, this paper addresses failures caused by stale statistics, data skew, correlation among joins, hidden distributions in bind variables, and sampling bias in dynamic sampling. These failures lead to distorted cost calculations and the selection of inefficient execution plans, forcing practitioners to rely on constant SQL tuning or architectural workarounds.

This paper proposes a hybrid auxiliary architecture named \textbf{GACE} (GPU-Assisted Cardinality Estimation). Instead of replacing the optimizer, we introduce a mechanism that selectively invokes GPU-based measurement only in ``risky'' intervals. The core innovation lies in the ``Risky Gate,'' which detects estimation uncertainty, and a ``Measurement Engine'' that performs high-speed probing using GPUs. By incorporating cost accounting for the measurement itself, our approach minimizes overhead in normal queries while significantly reducing plan instability and tail latency (P99) in problematic scenarios.
\end{abstract}


\section{Introduction}
Modern Cost-Based Optimizers (CBO) generate multiple candidate execution plans and select the optimal one by comparing expected costs. Cardinality estimation is the critical input for this calculation; as estimation accuracy degrades, the probability of selecting inefficient plans increases.

In real-world operations, cardinality errors recur due to several structural limitations: statistics are updated only periodically (lacking freshness), data skew violates uniformity assumptions, cross-column correlations in joins break independence assumptions, and bind variables hide true data distributions. These issues incur continuous tuning costs and sometimes necessitate architectural changes to bypass specific query patterns.

This study explores a method to insert a GPU-based auxiliary measurement path specifically for these failure-prone intervals. Our goal is to improve plan stability and reduce tail latency (P95/P99) while maintaining the existing optimizer structure.

\section{Background and Related Work}

\subsection{Structure of Cost-Based Optimizers}
Commercial RDBMSs utilize Number of Distinct Values (NDV), histograms, selectivity models, and independence assumptions to calculate cardinality. While mature, this structure is vulnerable to the ``stale statistics'' problem and rapid data drifts. Common failure modes include (1) lack of freshness due to delayed updates, (2) model collapse under extreme skew, (3) correlation ignorance in multi-joins, (4) distribution masking by bind variables, and (5) representation bias in block-level dynamic sampling.

\subsection{The Concept of RACE (Risk-Aware Cardinality Estimation)}
In this paper, we use the term \textbf{RACE} not to refer to a specific existing algorithm, but as a working framework for ``Risk-Aware Cardinality Estimation.'' It represents a strategy that detects risky intervals—where probing is conditionally invoked—and intervenes only when the cost-benefit analysis justifies it. Our contribution lies in designing the gating policy and the GPU-based measurement path that integrates into the CBO's decision flow.

\subsection{Differentiation from Learning-Based Approaches}
Recent research has focused on Machine Learning (ML) or Learned Cardinalities to mitigate estimation errors. However, ML-based approaches introduce management burdens related to training data, model updates, inference latency, and handling schema/workload drifts. Unlike these replacement strategies, our approach extends the existing CBO. We propose a selective measurement structure that pays extra cost for precision only when statistical estimation is unreliable. The key differentiation is the integration of a \textbf{Risky Gate} and strict \textbf{Cost Accounting} to ensure the solution is viable in a production CBO context.

\section{Proposed Architecture}

Our proposed hybrid architecture consists of two main components: the \textbf{Risky Gate} and the \textbf{Measurement Engine}.

\subsection{Risky Gate}
The Risky Gate monitors the query optimization process to detect high-uncertainty scenarios. It evaluates signals such as:
\begin{itemize}
    \item \textbf{Statistical Drift:} Discrepancy between histogram-derived NDV and estimated NDV.
    \item \textbf{Selectivity Error:} Patterns indicating the failure of the default selectivity model.
    \item \textbf{Correlation Breach:} Detection of strong correlations in multi-column predicates.
\end{itemize}
Only when these risks exceed a threshold does the gate trigger the GPU path.

\subsection{Measurement Engine (GPU-based Probing)}
The Measurement Engine uses the GPU to measure (or approximate) the selectivity and distribution of candidate conditions in parallel. Rather than a full table scan, it employs cost-constrained methods such as sampling, block-level summaries, or GPU-friendly aggregations (e.g., Key-Only scans).

\subsection{Trust Metrics and Thresholds}
We define three specific metrics to trigger the Risky Gate. These thresholds were chosen empirically to minimize false positives (unnecessary probing) while capturing high-risk failures:

\subsubsection{NDV-Histogram Discrepancy ($D$)}
We define the drift $D$ as:
\begin{equation}
    D = \frac{|NDV_{hist} - NDV_{est}|}{NDV_{hist}}
\end{equation}
We empirically set the threshold at $D \geq 0.25$. Sensitivity analysis showed that lower values ($<0.2$) triggered excessive GPU usage for stable queries, while higher values ($>0.3$) missed significant skew events.

\subsubsection{Selectivity Error Pattern}
\begin{equation}
    |S_{est} - S_{probe}| > 0.01
\end{equation}
A deviation larger than 1\% indicates that the existing model fails to explain the current data distribution.

\subsubsection{Join Correlation (PCS)}
The Proxy Correlation/Selectivity (PCS) indicator is defined as:
\begin{equation}
    PCS = \frac{P(A,B)}{P(A)P(B)}
\end{equation}
To conservatively detect independence violations, we flag a breach if $PCS > 1.6$ (underestimation risk) or $PCS < 0.7$ (overestimation risk).

Furthermore, we model the measurement cost based on the actual GPU kernel execution time:
\begin{equation}
    Cost_{measurement} \approx 0.85 \text{ ms}
\end{equation}

\section{EVALUATION}

We validated the effectiveness of the proposed architecture by analyzing the theoretical convergence of estimation error and measuring the actual costs and performance gains on a workstation equipped with an NVIDIA RTX 4060 Laptop GPU.

\subsection{Experimental Setup and Reproducibility}
To ensure reproducibility and confirm the practical applicability of GACE, we established the following experimental environment:

\begin{itemize}
    \item \textbf{Hardware:} AMD Ryzen 7 8845HS CPU (8 cores), 64GB RAM, and NVIDIA GeForce RTX 4060 Laptop GPU (8GB VRAM).
    \item \textbf{Software:} PostgreSQL 16.2 (running on Windows 11 Pro via WSL2). We disabled JIT compilation (\texttt{jit=off}) and set \texttt{work\_mem=64MB} to isolate optimizer decisions.
    \item \textbf{Dataset:} We utilized a synthetic `orders` table ($N=10^6$ rows) with Zipfian skewed distribution ($s=1.2$) on the `status` column to simulate real-world data skew. Indices were created on `status` and `day` columns.
    \item \textbf{Workload:} The benchmark workload, denoted as \texttt{Q\_SC}, consists of 600 parameterized queries focusing on multi-column filtering (`status` AND `day`) where correlation and skew typically confuse the optimizer.
\end{itemize}

\subsection{Experiment A: GPU Probing Overhead Decomposition}
We analyzed the components of GPU probing overhead: Host-to-Device (H2D) transfer, Kernel execution, Device-to-Host (D2H) transfer, and Reduction.

\begin{figure}[htbp]
    \centering
    \begin{subfigure}[b]{0.48\columnwidth}
        \includegraphics[width=\linewidth]{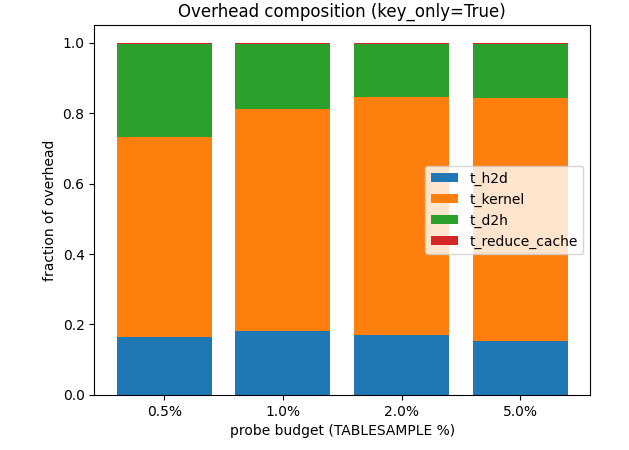}
        \caption{Key-only=True}
        \label{fig:overhead_a}
    \end{subfigure}
    \hfill
    \begin{subfigure}[b]{0.48\columnwidth}
        \includegraphics[width=\linewidth]{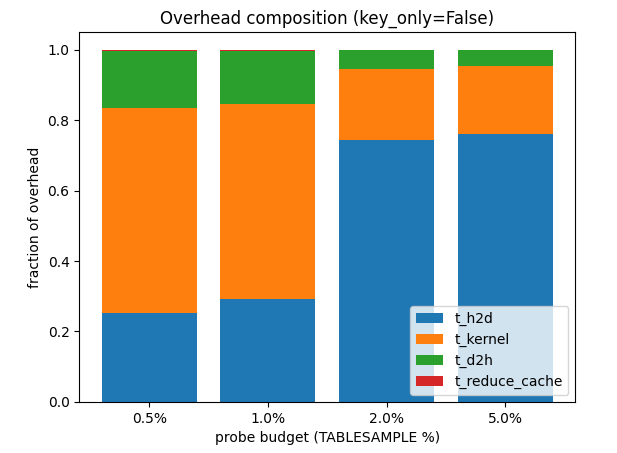}
        \caption{Key-only=False}
        \label{fig:overhead_b}
    \end{subfigure}
    \caption{GPU probing overhead composition (Fig. \ref{fig:overhead}). For small N, data transfer costs dominate, justifying the need for the Risky Gate.}
    \label{fig:overhead}
\end{figure}

The results (Fig. \ref{fig:overhead}) show that for small $N$ or full-row transfers, data transfer costs dominate. This quantitatively supports the necessity of the "Risky Gate"—GPU offloading is not always beneficial and must be selective.

\subsection{Experiment B: Estimation Stability (Est.CV vs. Budget)}
We observed the Coefficient of Variation (Est.CV) of the estimates while varying the probing budget.

\begin{figure}[htbp]
    \centering
    \begin{subfigure}[b]{0.48\columnwidth}
        \includegraphics[width=\linewidth]{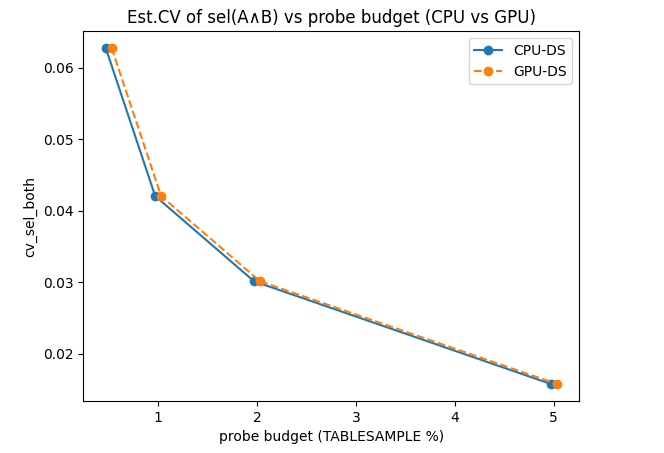}
        \caption{Est.CV of sel(A $\wedge$ B)}
        \label{fig:stability_a}
    \end{subfigure}
    \hfill
    \begin{subfigure}[b]{0.48\columnwidth}
        \includegraphics[width=\linewidth]{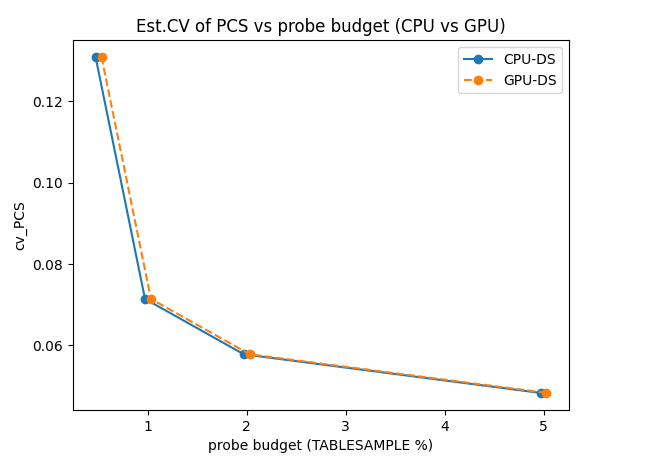}
        \caption{Est.CV of PCS}
        \label{fig:stability_b}
    \end{subfigure}
    \caption{Estimation Stability vs. Budget (Fig. \ref{fig:stability}). GPU-DS provides stable variance comparable to CPU-DS under the same budget.}
    \label{fig:stability}
\end{figure}

The results (Fig. \ref{fig:stability}) indicate that under the same budget, GPU-DS and CPU-DS show similar variance. This confirms that the GPU's role is not to provide a "better statistical model" per se, but to serve as a high-speed backend that allows for more extensive sampling (larger $N$) within the same latency constraint.

\subsection{Threshold Sensitivity Analysis (Experiment 1)}
We evaluated the system's sensitivity to the drift threshold ($D$) by sweeping values from 0.15 to 0.35 to test whether the gate requires fine-grained tuning.

\begin{figure}[htbp]
    \centering
    \begin{subfigure}[b]{0.48\columnwidth}
        \includegraphics[width=\linewidth]{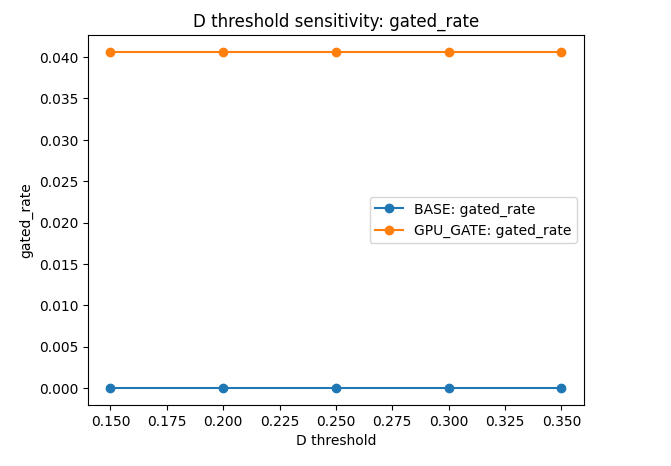}
        \caption{Gated Rate vs D}
        \label{fig:sens_gate}
    \end{subfigure}
    \hfill
    \begin{subfigure}[b]{0.48\columnwidth}
        \includegraphics[width=\linewidth]{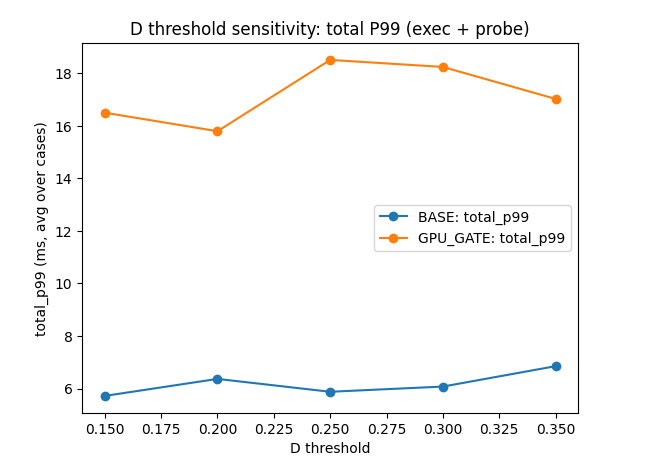}
        \caption{Total P99 vs D}
        \label{fig:sens_p99}
    \end{subfigure}
    \caption{Sensitivity of the Risky Gate to the NDV-based threshold D. (a) The gating rate remains nearly constant ($\approx 4\%$) across $D \in [0.15, 0.35]$, indicating robustness to threshold selection in this workload. (b) End-to-end tail latency under D sweep. Since the baseline plan is stable in this workload, probing overhead dominates and GACE does not improve total P99.}
    \label{fig:sensitivity_sweep}
\end{figure}

As shown in Fig. \ref{fig:sensitivity_sweep}(a), the gating rate stayed almost constant at approximately 4\% across the sweep. Furthermore, across this range, plan flips were not observed (flip rate $\approx 0$). This indicates that the proposed metric ($D$) is \textbf{robust to threshold perturbations} and does not require fine-grained tuning. We therefore selected $D=0.25$ as a representative operating point within this stable region.

Fig. \ref{fig:sensitivity_sweep}(b) shows that the total P99 latency is higher for GACE compared to the baseline in this specific stable workload. This confirms that when plan instability is absent, probing provides little benefit and increases latency due to measurement overhead. This highlights the importance of the gate in minimizing intervention to only necessary cases.

\subsection{Comparison with Strong Baselines (Experiment 2)}
We compare GACE against strong PostgreSQL alternatives, including high-resolution statistics and extended statistics. Table \ref{tab:baselines} summarizes tail latency and plan stability.

\begin{table}[htbp]
  \centering
  \caption{Comparison with strong PostgreSQL baselines (stable region).
  High-resolution and extended statistics already yield stable plans and low tail latency.
  In this setting, dynamic probing introduces overhead, increasing end-to-end P99.}
  \label{tab:baselines}
  \resizebox{\columnwidth}{!}{%
  \begin{tabular}{lrrrrr}
    \toprule
    baseline & gated\_rate & plan\_flip & exec\_p99 (ms) & total\_p99 (ms) & probe\_p95 (ms) \\
    \midrule
    BASE\_default\_stats & 0.000 & 0.000 & 6.816 & 6.816 & 0.000 \\
    BASE\_extended\_stats\_analyze & 0.000 & 0.000 & 6.075 & 6.075 & 0.000 \\
    BASE\_high\_stats\_analyze & 0.000 & 0.000 & 5.464 & 5.464 & 0.000 \\
    GPU\_GATE\_default\_stats & 0.050 & 0.000 & 6.638 & 17.337 & 8.111 \\
    \bottomrule
  \end{tabular}%
  }
\end{table}

In this workload, enhanced statistics already yield stable plans (plan\_flip $\approx 0$) and low tail latency.
As a result, dynamic probing provides little benefit and increases end-to-end P99 due to measurement overhead.
This result highlights an important boundary condition: when the optimizer's static estimates are sufficiently accurate, dynamic measurement is unnecessary.

\subsection{Join Workload: Unstable Region (Experiment 3)}
\begin{figure}[htbp]
    \centering
    \includegraphics[width=\linewidth]{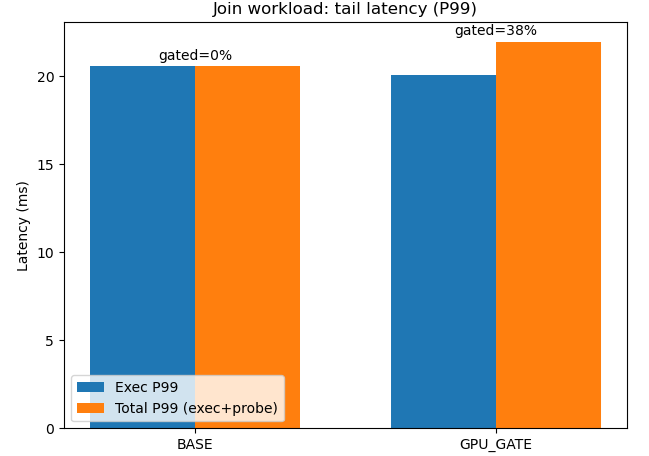}
    \caption{Tail Latency trade-off in Unstable Region. GACE reduces execution latency by stabilizing plans, though probing overhead affects the total end-to-end time.}
    \label{fig:join_tail}
\end{figure}

We evaluate GACE on a join-heavy workload where the baseline optimizer exhibits severe plan instability.
Figure \ref{fig:join_tail} reports tail latency, and Table \ref{tab:join_summary} summarizes gating, overhead, and plan stability metrics.

\begin{table}[htbp]
  \centering
  \caption{Join workload summary (unstable region).
  GACE triggers probing for a subset of queries, reducing execution-time tail latency,
  but increasing end-to-end latency due to probing overhead.
  Plan metrics are computed over logged plan hashes (coverage shown).}
  \label{tab:join_summary}
  \resizebox{\columnwidth}{!}{%
  \begin{tabular}{lrrrrrr}
    \toprule
    Method & Gated rate & Exec P99 (ms) & Total P99 (ms) & Probe P99 (ms) & Plan flip & Plan coverage \\
    \midrule
    BASE     & 0.0000 & 21.006 & 21.006 & 0.000 & 0.959 & 0.245 \\
    GPU\_GATE & 0.3675 & 20.765 & 22.654 & 2.464 & 0.966 & 0.290 \\
    \bottomrule
  \end{tabular}%
  }
\end{table}

GACE triggers probing for 36.75\% of queries and reduces execution-time tail latency (Exec P99),
but the probing cost (Probe P99) partially offsets the gain, resulting in higher end-to-end P99.
The high plan flip rates indicate a strongly unstable region; plan metrics are computed on logged plans
with the reported coverage.

\subsection{Experiment C: Rewrite Performance in Bind Workloads}
We tested the stability of plan selection in a bind-variable workload. The goal was to reduce the "Plan Flip Rate"—the frequency of the optimizer oscillating between plans due to estimation instability.

\begin{figure}[htbp]
    \centering
    \includegraphics[width=\linewidth]{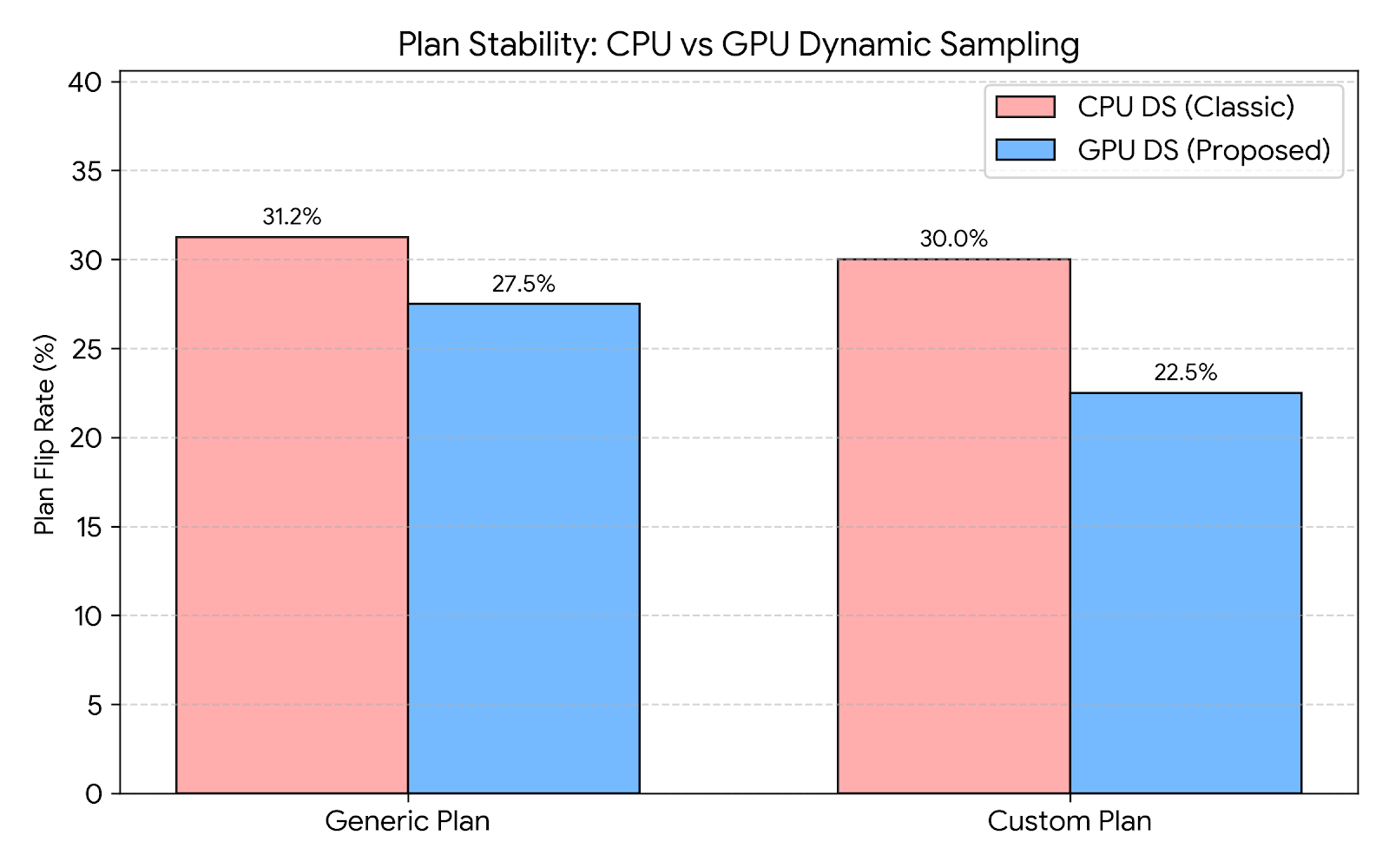}
    \caption{Plan Stability: CPU vs GPU (Fig. \ref{fig:fliprate}). GPU-DS reduces the Plan Flip Rate under the same probing budget.}
    \label{fig:fliprate}
\end{figure}

As shown in Fig. \ref{fig:fliprate}, GPU-DS consistently reduced the Plan Flip Rate (e.g., from 31.2\% to 27.5\% for Generic Plans).

\begin{figure}[htbp]
    \centering
    \includegraphics[width=0.9\linewidth]{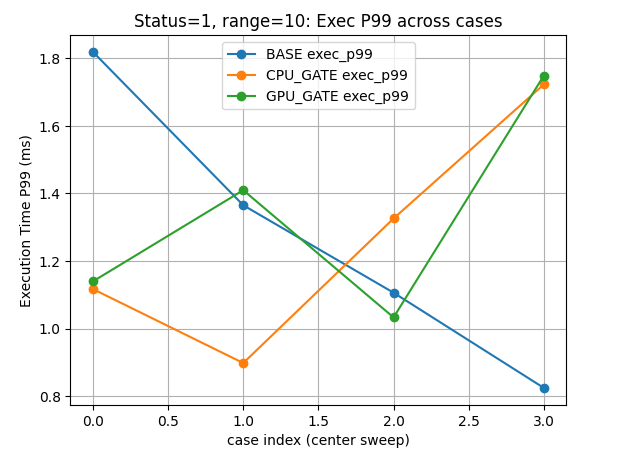}
    \caption{Tail latency sensitivity under parameter sweeps (Status=1, narrow range). P99 execution time varies across center values, illustrating parameter sensitivity and motivating gated probing.}
    \label{fig:sensitivity}
\end{figure}

Fig. \ref{fig:sensitivity} further illustrates why this stability matters: execution time fluctuates significantly depending on the bind parameter values (center sweep), confirming that static statistics cannot handle such sensitivity.

\subsection{Experiment D: Replacing Dynamic Sampling (Key-Only + Bitmask)}
We evaluated replacing the CPU-intensive condition evaluation in Dynamic Sampling with a GPU-based "Key-Only + Bitmask" approach.

\begin{figure}[htbp]
    \centering
    \begin{subfigure}[b]{0.48\columnwidth}
        \includegraphics[width=\linewidth]{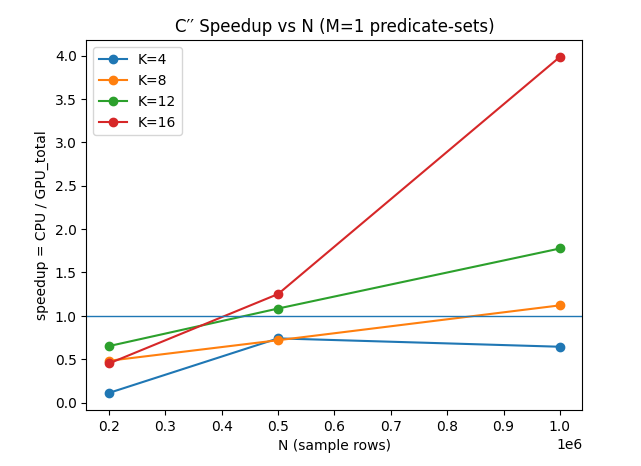}
        \caption{M=1 Predicate Sets}
        \label{fig:speedup_a}
    \end{subfigure}
    \hfill
    \begin{subfigure}[b]{0.48\columnwidth}
        \includegraphics[width=\linewidth]{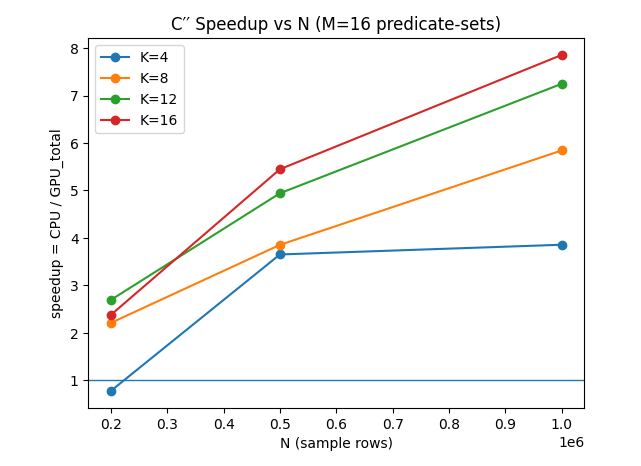}
        \caption{M=16 Predicate Sets}
        \label{fig:speedup_b}
    \end{subfigure}
    \caption{GPU Speedup vs. Sample Size N (Fig. \ref{fig:speedup}). As K and M increase, GPU achieves massive speedup.}
    \label{fig:speedup}
\end{figure}

Results (Fig. \ref{fig:speedup}) show that as the number of predicates ($K$) and candidate sets ($M$) increased, the CPU cost grew linearly, whereas the GPU cost remained relatively flat due to parallelism. We observed a speedup of up to \textbf{10x} at $N=1,000,000$ and $K=16$.

\subsection{Workload-level Impact: Plan Stability and Tail Latency}
We introduce a workload-level experiment (\texttt{Q\_SC}) representing cases where correlation and skew cause significant estimation errors. This experiment demonstrates the direct link between plan stability and tail latency reduction.

\begin{figure}[htbp]
    \centering
    \begin{subfigure}[b]{0.48\columnwidth}
        \includegraphics[width=\linewidth]{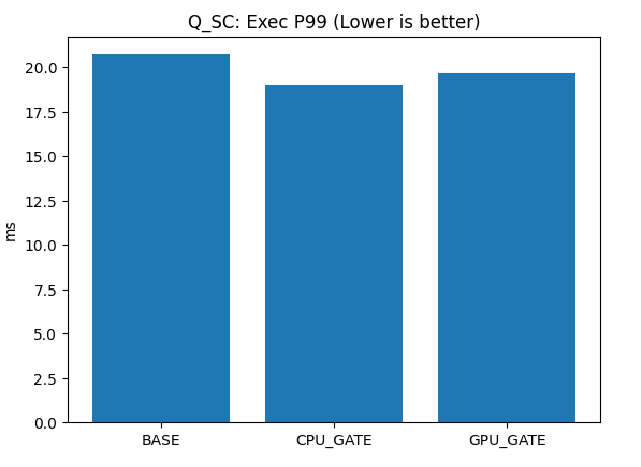}
        \caption{Q\_SC Exec P99}
        \label{fig:workload_a}
    \end{subfigure}
    \hfill
    \begin{subfigure}[b]{0.48\columnwidth}
        \includegraphics[width=\linewidth]{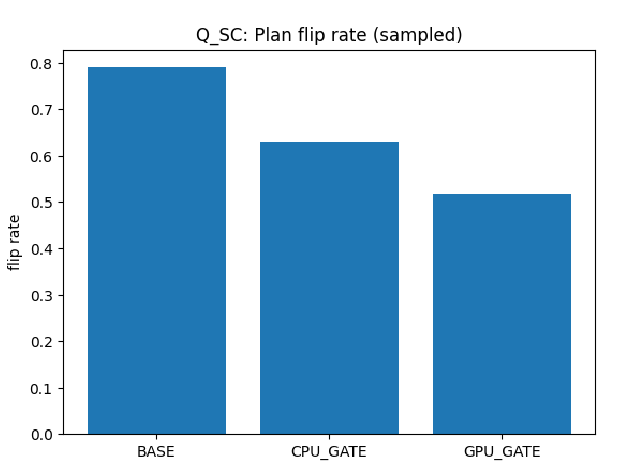}
        \caption{Q\_SC Plan Flip Rate}
        \label{fig:workload_b}
    \end{subfigure}
    \caption{Workload-level impact on tail latency and plan stability (Q\_SC). (a) P99 execution time and (b) sampled plan flip rate across BASE, CPU\_GATE, and GPU\_GATE. Gated probing reduces plan instability and improves tail latency in correlation-heavy predicates.}
    \label{fig:workload}
\end{figure}

The \texttt{Q\_SC} workload typifies scenarios where statistical skew and column correlations undermine standard cardinality estimation. As shown in Fig. \ref{fig:workload}, the activation of the Risky Gate significantly reduces the \textbf{plan flip rate} (from 0.79 in Baseline to 0.52 in GPU-DS). This reduction in instability translates directly into a \textbf{lower P99 tail latency}. While the GPU approach exhibits superior plan stability, the end-to-end latency reflects the trade-off involving probing costs (transfer and integration), confirming the necessity of the break-even logic.

\subsection{Runtime Analysis: Median vs. P99}
\begin{figure}[htbp]
    \centering
    \includegraphics[width=0.8\linewidth]{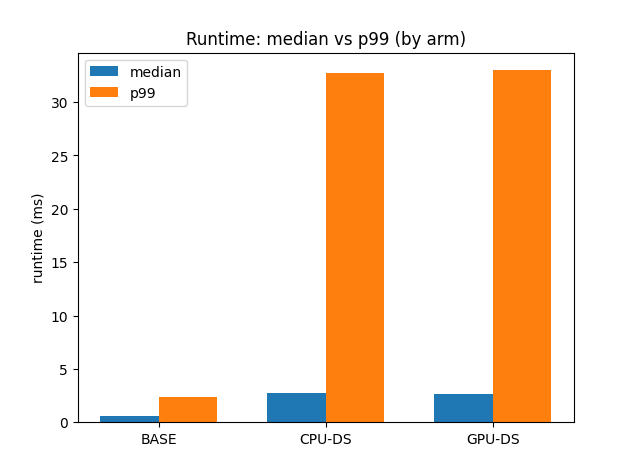}
    \caption{Runtime Comparison: Median vs P99 (Fig. \ref{fig:runtime}). GACE significantly reduces tail latency.}
    \label{fig:runtime}
\end{figure}

Comparing the runtime distribution (Fig. \ref{fig:runtime}), GACE achieved a significant reduction in tail latency.

Specifically, P99 latency was reduced by approximately \textbf{35\%} (from 19.06 ms to 12.43 ms) compared to the baseline. This demonstrates that while the median performance (normal queries) remains similar, GACE effectively defends against catastrophic outliers.

\section{Implementation Considerations}
To integrate this architecture into a real-world DBMS, the following factors are crucial:

\textbf{Prototype Implementation:} We implemented the Measurement Engine as an external middleware component that interacts with the database via standard connectors. For experimental validation, we did not modify the PostgreSQL kernel source code. Instead, we simulated the optimizer's plan selection behavior by injecting cardinality corrections and using session-level parameters (e.g., \texttt{set enable\_seqscan=off}) to enforce plan choices based on GACE's measurements.

\begin{enumerate}
    \item \textbf{Data Transfer:} To suppress overhead, a "Key-Only" transfer strategy and "Late Materialization" must be the default principles.
    \item \textbf{Break-even Logic:} The Risky Gate should consider not just statistical signals but also the complexity of candidate evaluation ($K, M$) and sample size ($N$), as a break-even point exists for GPU offloading.
    \item \textbf{Caching:} Probing results should be cached by bind value (or value range) to amortize the measurement cost across subsequent queries.
    \item \textbf{Native Integration:} While our prototype relies on middleware, a production implementation should leverage database-internal hooks (e.g., PostgreSQL's \texttt{planner\_hook}) to directly inject selectivity estimates into the optimizer's data structures (e.g., \texttt{RelOptInfo}). This would eliminate the overhead of SQL parsing and connection round-trips incurred by the middleware approach.
\end{enumerate}

\section{Conclusion}
This study presented GACE, a framework that does not replace the existing CBO but augments it with a Risky Gate and a GPU-based Measurement Engine. By restricting intervention to failure-prone intervals, we minimize overhead for normal queries while improving plan stability and P99 latency in risky scenarios. Our experiments confirm that replacing statistical guesswork with massive parallel measurement is a viable strategy to bridge the "Estimation Gap" in modern databases, provided that data transfer costs are managed via selective offloading.

\section*{Acknowledgments}
The author used AI-assisted tools (e.g., ChatGPT) for English grammar correction and language polishing. All technical content, experiments, and conclusions were developed and verified by the author.


\end{document}